\documentstyle[aps,epsf,twocolumn]{revtex}
\begin{document}
\title{Pauli Tomography: complete characterization of a single qubit device}
\author{F. De Martini*, Giacomo Mauro D'Ariano$^{\dagger }$, Andrea Mazzei* \& Marco
Ricci*}
\address{*Istituto Nazionale di Fisica della Materia, Dipartimento di Fisica\\
Universit\`{a} ''La Sapienza'', Roma, 00185 Italy,\\
$^{\dagger }$Istituto Nazionale per la Fisica della Materia and Dipartimento%
\\
di Fisica ''Alessandro Volta''\\
Universit\`{a} di Pavia, Via Bassi, 6, I-27100 Pavia, Italy}
\maketitle

\begin{abstract}
The marriage of Quantum Physics and Information Technology, originally
motivated by the need for miniaturization, has recently opened the way to
the realization of radically new information-processing devices, with the
possibility of guaranteed secure cryptographic communications, and
tremendous speedups of some complex computational tasks. Among the many
problems posed by the new information technology [I.L. Chuang \& M.A.
Nielsen, {\it Quantum Information and Quantum Computation}, (Cambridge Univ.
Press, Cambridge, 2000).] there is the need of characterizing the new
quantum devices, making a complete identification and characterization of
their functioning. As we will see, quantum mechanics provides us with a
powerful tool to achieve the task easily and efficiently: this tools is the
so called quantum entanglement, the basis of the quantum parallelism of the
future computers. We present here the first full experimental quantum
characterization of a single-qubit device. The new method, we may refer to
as ''quantum radiography'', uses a Pauli Quantum Tomography at the output of
the device, and needs only a single entangled state at the input, which
works on the test channel as all possible input states in quantum parallel.
The method can be easily extended to any n-qubits device.
\end{abstract}

How do we usually characterize the operation of a device? Actually, we are
interested just in linear devices, since quantum dynamics is intrinsecally
linear. Any linear device, either quantum or classical (examples are: an
optical lens or a good amplifier), can be completely described by a {\it %
transfer matrix} which gives the output vector by matrix-multiplying the
input vector. In quantum mechanics the inputs are density operators $\rho
_{in}$ and the role of the transfer matrix is played by the so called {\it %
quantum operation} \cite{2} of the device, that here we will denote by {\it E%
}. Thus the output state $\rho _{out}$\ is given by the quantum operation
applied to the input state as follows:
\begin{equation}
\rho _{out}=\frac{E\left( \rho _{in}\right) }{Tr\left[ E\left( \rho
_{in}\right) \right] }
\end{equation}
and the normalization constant $Tr\left[ E\left( \rho _{in}\right) \right] $%
\ is also the probability of occurence of the transformation {\it E}, when
there are other possible alternatives, such as when we consider the state
transformation due to a measuring device for a given outcome.

Now the problem is: how to reconstruct the form of E experimentally? One
would be tempted to adopt the conventional method \cite{1} of running a {\it %
basis} of all possible inputs, and measuring the corresponding outputs by
{\it quantum \ tomography} \cite{3}. However, since the states $\rho $ are
actually operators, not vectors, in order to get all possible matrix
elements we would need to run a complete orthogonal basis of quantum states $%
\left| n\right\rangle $\ along with their linear combination $2^{\frac{1}{2}%
}\left( \left| n^{\prime }\right\rangle +i^{k}\left| n^{\prime \prime
}\right\rangle \right) $, with k=0,1,2,3 and i denoting the imaginary unit
(this is a simple consequence of the polarization identity). However, the
availability of such a set of states in the laboratory is, by itself, a very
hard technological problem (states with a precise varying number of photons
and, even worst, their superposition, are still a dream for
experimentalists).

The quantum parallelism intrinsic of entanglement now comes to help us,
running all possible input states in parallel by using only a single
entangled state as the input! This was first shown in \cite{4}. Hence, we
don't need to prepare a complete set of states, but just one entangled
state, a state commonly available in modern quantum optical laboratories!

\section{Pauli Tomography}

Assume for semplicity, and with no loss of generality, that the entangled
state spans a 4-dimensional Hilbert space $H_{1}\otimes H_{2}$, each space H$%
_{i}$ of dimension d=2. The input entangled state,
\begin{equation}
\left| \left. \Psi \right\rangle \right\rangle =\sum_{nm}\Psi _{nm}\left|
nm\right\rangle
\end{equation}
expressed in terms of the basis vectors $\left| n\right\rangle \otimes
\left| m\right\rangle =\left| nm\right\rangle $ of the two spaces H$_{i}$ (i
=1,2) is used as shown in Fig. 1 where the entangled systems consist of two
single-mode optical beams. The key feature of the method implies that only
one of the two systems, say system i=1, is input into the {\it unknown}
device E, whereas the other is left untouched. This setup leads to the
output state R$_{out}$, which in tensor notation writes as follows
\begin{equation}
R_{out}=E\otimes I\left| \left. \Psi \right\rangle \right\rangle
\left\langle \left\langle \Psi \right. \right|
\end{equation}
where I denotes the identical operation. It is a result of linear algebra
that R$_{out}$ is in one-to-one correspondence with the quantum operation E,
as long as the state $\left| \left. \Psi \right\rangle \right\rangle $ is
full-rank, i. e. it has non-vanishing components on the whole state-space of
each system, such as, for instance, a so called maximally entangled state.
Full-rank entangled states can be easily generated by Spontaneous Parametric
Down Conversion of the vacuum state, as in the experiment reported here.
Note that by this method the problem of availability of all possible input
states is solved: we just need a single entangled state $\left| \left. \Psi
\right\rangle \right\rangle $, which works as all possible inputs in a sort
of quantum parallelism!

Now, how to characterize the entangled state R$_{out}$ at the output? We
obviously need to perform many measuremeuts on an ensemble of equally
prepared quantum systems, since, due to the {\it no-cloning theorem} \cite{5}
we cannot determine the state of a single system \cite{6}. For this purpose
a technique for the full determination of the quantum state has been
introduced and developed since 1994. The method named {\it Quantum Tomography%
} \cite{3} has been initially introduced for the state of a single-mode of
radiation, the so called {\it Homodyne Tomography}, and thereafter it has
been generalized to any quantum system. The basis of the method is just
performing measurements of a suitably complete set of observables called
{\it quorum}. For our needs, we just have to measure jointly a quorum of
observables, here the spin observables $\sigma _{i}$ (i=0,1,2,3), on the two
entangled systems at the output, in order to determine the output state $%
R_{out}$, and hence the quantum operation E.

In this paper we present the first complete experimental characterization of
a quantum device, which in our case will be a single-qubit device. The qubit
is encoded on polarization of single photons in the following way
\begin{equation}
\begin{array}{ccc}
\left| 0\right\rangle =\left| 1\right\rangle _{h}\left| 0\right\rangle _{v}
& , & \left| 1\right\rangle =\left| 0\right\rangle _{h}\left| 1\right\rangle
_{v}
\end{array}
\end{equation}
namely with the ''logical zero'' state corresponding to a single
horizontally polarized photon and the ''logical one'' state corresponding to
a single vertically polarized photon. In the following we will denote by h
and v the annihilation operators of the horizontally and vertically
polarized modes of radiation associated to a fixed wave-vector, {\bf k}.
Using single photon states we encode a qubit on the polarization. In the
polarization representation, the Pauli matrices write as follows:
\begin{equation}
\begin{array}{c}
\sigma _{1}=h^{\dagger }v+v^{\dagger }h \\
\sigma _{2}=i\left( h^{\dagger }v-v^{\dagger }h\right) \\
\sigma _{3}=h^{\dagger }h-v^{\dagger }v
\end{array}
\end{equation}
The ring of Pauli matrices is completed by including the identity $\sigma
_{0}=h^{\dagger }h+v^{\dagger }v$. In the following we will denote by $\vec{%
\sigma}$\ the column three-vector of operators $\vec{\sigma}$ = $\left(
\sigma _{1},\sigma _{2},\sigma _{3}\right) $, and by $\sigma $ the column
tetra-vector $\sigma $ = $\left( \sigma _{0},\sigma _{1},\sigma _{2},\sigma
_{3}\right) $, and use Greek indices for three-vectors components $\alpha $
=1, 2, 3 (or $\alpha $ = x, y, z), and Latin indices for tetra-vector
components: i = 0, 1, 2, 3.

A wave-plate changes the two radiation modes according to the matrix
transformation:
\begin{equation}
\left(
\begin{array}{c}
h \\
v
\end{array}
\right) \longrightarrow w\left( \phi ,\theta \right) ^{\dagger }\left(
\begin{array}{c}
h \\
v
\end{array}
\right) w\left( \phi ,\theta \right) \equiv W\left( \phi ,\theta \right)
\left(
\begin{array}{c}
h \\
v
\end{array}
\right)
\end{equation}
where the matrix W$\left( \phi ,\theta \right) $ is given by:
\begin{eqnarray}
W\left( \phi ,\theta \right) &=&\left(
\begin{array}{cc}
\cos \theta & -\sin \theta \\
\sin \theta & \cos \theta
\end{array}
\right) \left(
\begin{array}{cc}
1 & 0 \\
0 & e^{i\phi }
\end{array}
\right) \left(
\begin{array}{cc}
\cos \theta & \sin \theta \\
-\sin \theta & \cos \theta
\end{array}
\right)  \nonumber \\
&=&\left(
\begin{array}{cc}
z_{+}+cz_{-} & sz_{-} \\
sz_{-} & z_{+}-cz_{-}
\end{array}
\right)
\end{eqnarray}
where $s=\sin 2\theta ,$ $c=\cos 2\theta ,$ $\theta $ is the wave-plate
orientation angle around the wave-vector {\bf k}, $z_{\pm }=\frac{1}{2}%
\left( 1\pm e^{i\phi }\right) ,$ $\phi =\frac{2\pi \delta }{\lambda },$ $%
\lambda $ is the wave-lenght and $\delta $\ is the optical path through the
plate. Special cases are the $\frac{\lambda }{4}$\ plate which can be used
with $\theta =\frac{\pi }{4}$\ to give the right and leftg circularly
polarized {\it modes}
\begin{equation}
\left(
\begin{array}{c}
r \\
l
\end{array}
\right) =W\left( \frac{\pi }{2},\frac{\pi }{4}\right) \left(
\begin{array}{c}
h \\
v
\end{array}
\right) =\frac{e^{i\pi /4}}{\sqrt{2}}\left(
\begin{array}{c}
h+iv \\
-ih+v
\end{array}
\right)
\end{equation}
and the $\frac{\lambda }{2}$\ plate wich can be used to give the diagonal
linearly polarized {\it modes}
\begin{equation}
\left(
\begin{array}{c}
a \\
b
\end{array}
\right) =W\left( \pi ,\frac{\pi }{8}\right) \left(
\begin{array}{c}
h \\
v
\end{array}
\right) =\frac{1}{\sqrt{2}}\left(
\begin{array}{c}
h+v \\
h-v
\end{array}
\right) .
\end{equation}
The Heisenberg picture evolution of the Pauli matrices corresponding to the
unitary {\it U} on the qubit Hilbert space will be given by a rotation,
which we will denote as follows
\begin{equation}
U^{\dagger }\vec{\sigma}U=R(U)\vec{\sigma}.  \label{10}
\end{equation}
In particular, for a $\phi -$wave-plate we have the rotation matrix
\begin{equation}
R\left( W\left( \phi ,\theta \right) \right) =\left(
\begin{array}{ccc}
s^{2}+c^{2}\cos \phi & -c\cos \phi & sc(1-\cos \phi ) \\
c\sin \phi & \cos \phi & -s\sin \phi \\
sc(1-\cos \phi ) & s\sin \phi & c^{2}+s^{2}\cos \phi
\end{array}
\right)  \label{matrixtrans}
\end{equation}
In particular, for a $\frac{\lambda }{2}$-wave-plate we have
\begin{eqnarray}
W\left( \pi ,\theta \right) &=&\left(
\begin{array}{cc}
\cos 2\theta & \sin 2\theta \\
\sin 2\theta & -\cos 2\theta
\end{array}
\right)  \nonumber \\
R\left( \pi ,\theta \right) &=&\left(
\begin{array}{ccc}
-\cos 4\theta & 0 & \sin 4\theta \\
0 & -1 & 0 \\
\sin 4\theta & 0 & \cos 4\theta
\end{array}
\right)
\end{eqnarray}
The $\sigma _{z}$-photo-detector is achieved as in Fig 2. From Eq. \ref
{matrixtrans} we can see that a $\sigma _{x}$-detector can be obtained by
preceding the $\sigma _{z}$ -detector with a $\lambda $/2-wave-p1ate
oriented at $\theta $= $\pi $/8, whereas a $\sigma _{y}$-detector is
obtained by preceding the $\sigma _{z}$-detector with a $\lambda $%
/4-wave-plate oriented at $\theta $= $\pi $/4. When collecting data at a $%
\sigma _{\alpha }$-detector, we will denote by $\sigma _{\alpha }=\pm $1 the
random outcome, with $\sigma _{\alpha }=\pm 1$ corresponding to the
h-detector flashing, and $\sigma _{\alpha }=-1$ corresponding to the
v-detector flashing instead. The experimental averages for the complete
setup must coincide with the following theoretical expectation values
\begin{equation}
\overline{s_{i}^{(1)}s_{j}^{(2)}}=\left\langle \left\langle \Psi \right.
\right| (U^{\dagger }\otimes I)(\sigma _{i}^{(1)}\otimes \sigma
_{j}^{(2)})(U\otimes I)\left| \left. \Psi \right\rangle \right\rangle
\end{equation}
and, in particular, $\overline{s_{i}^{(1)}}\equiv \overline{%
s_{i}^{(1)}s_{0}^{(2)}}$ and $\overline{s_{i}^{(2)}}\equiv \overline{%
s_{0}^{(1)}s_{2}^{(2)}}$ now s$_{i}^{(n)}$denoting the random outcome of the
detector of the {\it n}th beam (n=1,2) in the entangled state. For maximally
entantangled states we have also $\overline{s_{\alpha }^{(1)}}=$ $\overline{%
s_{\alpha }^{(2)}}=0$ for all $\alpha =x,y,z$. The theoretical expectations
for the setup without the device to be characterized are be given by:
\begin{equation}
\left\langle \left\langle \Psi \right. \right| (\sigma _{i}^{(1)}\otimes
\sigma _{j}^{(2)})\left| \left. \Psi \right\rangle \right\rangle =Tr\left[
\Psi ^{+}\sigma _{i}\Psi \sigma _{j}^{\ast }\right]  \label{14}
\end{equation}
Where $\Psi $ denotes the matrix of the state on the customary basis of
eigenvectors of $\sigma _{z}$. In particular, for the four Bell states
\begin{equation}
\frac{1}{\sqrt{2}}\left| \left. \sigma _{j}\right\rangle \right\rangle
=\sigma _{j}\otimes I\frac{1}{\sqrt{2}}\left| \left. I\right\rangle
\right\rangle
\end{equation}
we have
\begin{equation}
\frac{1}{\sqrt{2}}\Delta _{ij}\left( \sigma _{k}\right) =\delta
_{ij}H_{kj},\;\;\;\;\;\;\;H=\left[
\begin{array}{cccc}
1 & 1 & -1 & 1 \\
1 & 1 & 1 & -1 \\
1 & -1 & -1 & 1 \\
1 & -1 & 1 & 1
\end{array}
\right]
\end{equation}
Using Eq. \ref{10} and \ref{14} we obtain
\begin{equation}
\overline{s_{\alpha }^{1}s_{\beta }^{2}}=\sum_{\lambda }R_{\alpha \lambda
}(U)\Delta _{\lambda \beta }(\Psi )
\end{equation}
In particular, in the lab we use the multiplet state corresponding to $\Psi
=\sigma _{x}/\sqrt{2}$, whence we have
\[
\overline{s_{\alpha }^{1}s_{\beta }^{2}}=R_{\alpha \beta }(U)\Delta _{1\beta
}(\Psi )
\]

\section{Method, Apparatus and Results}

The input maximally entangled state $\left| \left. \Psi \right\rangle
\right\rangle $, expressed by Eq 2, was SPDC generated in the laboratory by
an optical parametric amplifier (OPA) physically consis-ting of a nonlinear
(NL) BBO (\ss -barium-borate) crystal plate, 2 mm thick, cut for Type II
phase matching and excited by a pulsed mode-locked ultraviolet laser UV
having pulse duration $\tau =140$ fsec and wavelength (wl) $\lambda _{p}$%
=397.5 nm. Precisely, the apparatus was set to generate on the two modes
{\bf k}$_{i}$, i.e. the entangled systems i =1,2, single photon couples in a
polarization entangled ''triplet'' state, viz. $\left| \left. \Psi
\right\rangle \right\rangle =2^{-1/2}\left| \left. \sigma _{x}\right\rangle
\right\rangle $, according to Eq. 15. The wl of the emitted photons was $%
\lambda $= 795 nm. The measurement apparatus consisted of two equal
polarizing beam splitters PBS$_{i}$ with output modes coupled to four equal
Si-avalanche photo-detectors SPCM-AQR14 with quantum efficiencies QE $\simeq
$ 0.42. The beams exciting the detectors were filtered by equal interference
filters within a bandwidth $\Delta \lambda $ = 6 nm. The detector output
signals were finally analyzed by a computer. We want now to determine
experimentally by this apparatus the matrix elements of the state $\left|
\left. \Psi \right\rangle \right\rangle $ expressed by Equation 2. This can
be achieved as follows: from the trivial identity
\begin{equation}
\left\langle nm\right| \left. \left. \Psi \right\rangle \right\rangle =\Psi
_{nm}
\end{equation}
we obtain the matrix $\Psi _{nm}$ for the {\it input states} in terms of the
following ensemble averages
\begin{equation}
\Psi _{nm}=e^{i\varphi }\frac{\left\langle \left\langle \Psi \right. \right|
\left. 01\right\rangle \left\langle nm\right| \left. \left. \Psi
\right\rangle \right\rangle }{\sqrt{\left\langle \left\langle \Psi \right.
\right| \left. 01\right\rangle \left\langle 01\right| \left. \left. \Psi
\right\rangle \right\rangle }}  \label{20}
\end{equation}
where the unmeasurable phase factor is given by: $\exp \left( i\varphi
\right) =\Psi _{01}/\left| \Psi _{01}\right| .$ The choice of the vector $%
\left| 01\right\rangle $ is arbitrary as it is needed only for the sake of
normalization: e.g. we could have used $\left| 10\right\rangle $ or $\left|
11\right\rangle $, instead. Using the {\it tomographic expansion} over the
four Pauli matrices \cite{3},\cite{4} we see that, in virtue of Eq. \ref{20}%
, the matrix element of the input state is obtained from the following
experimental averages:
\begin{equation}
\Psi _{nm}=\frac{1}{4\sqrt{p}}\sum_{ij}Q_{ij}(nm)\overline{%
s_{i}^{(1)}s_{j}^{(2)}}
\end{equation}
where
\begin{equation}
p=\left\langle \left\langle \Psi \right. \right| \left. 01\right\rangle
\left\langle 01\right| \left. \left. \Psi \right\rangle \right\rangle =\frac{%
1}{4}\left( 1+s_{3}^{(1)}\right) \left( 1-s_{3}^{(2)}\right)  \label{23}
\end{equation}
is the fraction of coincidences with both $\sigma _{z}$-detectors firing on
h, and the matrix Q(nm) is given by
\begin{equation}
Q_{ij}(nm)=\left\langle n\right| \sigma _{i}\left| 0\right\rangle
\left\langle m\right| \sigma _{j}\left| 1\right\rangle  \label{24}
\end{equation}
the unitary matrix $U_{nm}$ of the device is now obtained with the same
averaging above but now for the state at the output of the device: $\left|
\left. U\Psi \right\rangle \right\rangle =\left( U\otimes I\right) \left|
\left. \Psi \right\rangle \right\rangle .$ Therefore, we now have:
\begin{equation}
\left( U\Psi \right) _{nm}=e^{i\varphi }\frac{\left\langle \left\langle
U\Psi \right. \right| \left. 01\right\rangle \left\langle nm\right| \left.
\left. \Psi U\right\rangle \right\rangle }{\sqrt{\left\langle \left\langle
U\Psi \right. \right| \left. 01\right\rangle \left\langle 01\right| \left.
\left. \Psi U\right\rangle \right\rangle }}  \label{25}
\end{equation}
\ where we use again Eqns.\ref{23}, \ref{24}, but now the average expressed
by 20 is carried out over the output state $\left| \left. U\Psi
\right\rangle \right\rangle $. The (complex) parameters U$_{nm}$ are
obtained from Eq. \ref{25} by matrix inversion. This is of course possible
since the matrix $\Psi $ is invertible, in virtue of the entangled character
of $\left| \left. \Psi \right\rangle \right\rangle $.

The experimental demonstration of the tomographic process is given in
Figures 3 and 4 where both {\it real} and {\it imaginary parts} of the four
components of the matrix U are reported for two different ''unknown''
devices inserted in the mode {\bf k}$_{1}$ of the tomographic apparatus. The
experimental results are shown together with the corresponding data
evaluated theoretically. Furthermore the experimental ''variance'' of the
data are also reported. Each ''unknown'' device is represented in one
figure, namely:

\underline{Figure 3}: a single Waveplate $\left[ \varphi =0.45\pi ;\theta
=-0.138\pi \right] $ i.e., with retardation phase $\varphi $ = $\left(
0.45\pi \right) $ and orientation angle respect to the ''horizontal''
direction ''h'' : $\theta $ = $\left( -0.138\pi \right) $.

\underline{Figure 4}: a combination of 2 Waveplates: a Waveplate $\left[
\varphi =0.45\pi ;\theta =-0.138\pi \right] $ followed by a $\lambda $/2
Waveplate $\left[ \varphi =\pi ;\theta =+0.29\pi \right] $.

As we may see, the experimental results are found in good agreement with
theory.

\section{Conclusions}

We have given the first demonstration, in a simple single-qubit context, of
a novel Tomographic method which is able to fully characterize the
properties of any device acting on a quantum system by exploiting for the
first time the complete intrinsic parallelism of the quantum entanglement.
This method establishes a new fundamental framework of utterly paradigmatic
relevance in the domains of modern Quantum Measurement theory and Quantum
Information.

Our method is expected to be of general and far reaching relevance. In
facts, it can be adopted within more general and complex multi-qubit
systems. For instance by this method a full characterization of a two-qubits
device, such as a controlled-NOT, can be achieved. In this case we just need
to double the input and the measurement setup, by providing two input
entangled states and four detectors coupled at the outputs of the two pairs
of the device channels. The full quantum characterization of the the device
is finally obtained by a joint tomographic reconstruction on both channels
of the device.

We are indebted with the FET European Network on Quantum Information and
Communication (Contract IST-2000-29681-ATESIT), the I.N.F.M. PRA 2001
''clon'' and with M.U.R.S.T. for funding.

\centerline{\bf Figure Captions}

\vskip 8mm

\parindent=0pt

\parskip=3mm

FIG. 1. General experimental scheme of the method for the tomographic
estimation of the quantum operation of a single qubit device. Two identical
quantum systems, e.g. two optical beams as in the present experiment, are
prepared in an entangled state $\left. \left| \Psi \right\rangle
\right\rangle $. One of the systems undergoes the quantum operation E,
whereas the other is left untouched. At the output one makes a quantum
tomographic estimation, by measuring jointly two observa-bles from a {\it %
quorum} \{O(l)\}. In the present experiment the quorum is represented by the
set of Pauli operators.

FIG. 2. Pauli-matrix measurement apparatus for photon polarization qubits
inserted at the end of each test optical beam.

FIG. 3. Experimental characterization by Pauli Tomography of a single
optical Waveplate $\left[ \varphi =0.45\pi ;\theta =-0.138\pi \right] $
inserted on channel {\bf k}$_{1}$ with the following optical properties:
retardation phase $\varphi =\left( 0.45\pi \right) $; orientation angle of
the optical axis respect to the laboratory horizontal direction ''h'': $%
\theta =\left( -0.138\pi \right) $. The experimental real and imaginary
parts of the four matrix elements U$_{ij}$ of the Waveplate are shown
together with the related measured statistical variances. The corresponding
theoretical values are shown for comparison.

FIG. 4. Experimental Chracterization by Pauli Tomography of a combination of
two optical Waveplates: the Waveplate $\left[ \varphi =0.45\pi ;\theta
=-0.138\pi \right] $ (cfr.Fig.3) followed by a $\lambda $/2 Waveplate $\left[
\varphi =\pi ;\theta =+0.29\pi \right] $. The experimental real and
imaginary parts of the four matrix elements U$_{ij}$ of the combination are
shown together with the related measured statistical variances. The
corresponding theoretical values are shown for comparison \newline

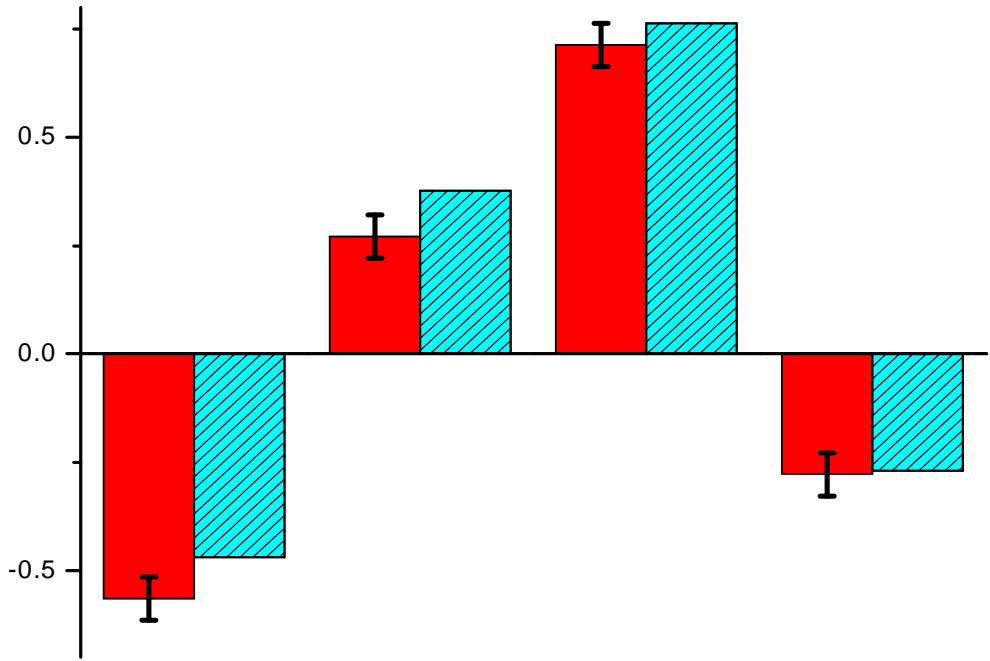
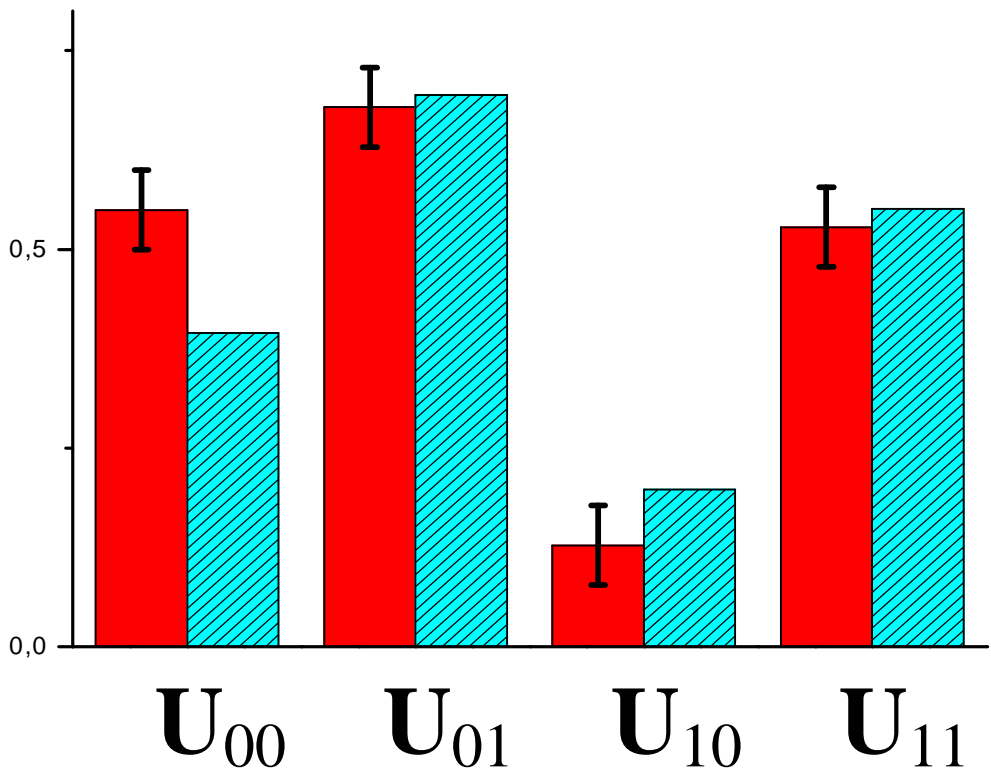
\end{document}